\documentclass[
  aip,
  pop,
  amsmath,amssymb,
  reprint,
]{revtex4-1}
\usepackage{graphicx,color}
\usepackage{amsmath}
\usepackage{natbib}
\usepackage{epsfig}
\begin{document}

\title{Onset of negative dispersion in one-component-plasma revisited}

\author{Sergey A. Khrapak}
\affiliation{Aix Marseille University, CNRS, Laboratoire PIIM, Marseille, France \\
Forschungsgruppe Komplexe Plasmen, Deutsches Zentrum f\"{u}r Luft- und Raumfahrt (DLR),
Oberpfaffenhofen, Germany \\
Joint Institute for High Temperatures, Russian Academy of Sciences, Moscow, Russia}

\date{\today}

\begin{abstract}
A simple approach to describe the long-wavelength dispersion of the longitudinal (plasmon) mode of the classical one-component-plasma (OCP), with the main objective to correctly capture the onset of negative dispersion, is proposed. The approach is applicable to both three-dimensional and two-dimensional OCP. The predicted onset of negative dispersion compares well with the available results from numerical simulations and more sophisticated theoretical models.   
\end{abstract}

\pacs{52.27.Aj, 52.27.Gr, 62.60.+v}
\maketitle

The one-component-plasma (OCP) represents an idealized physical system of identical point-like charges immersed in a uniform neutralizing background of opposite charge.~\cite{BrushJCP1966,BausPR1980,IchimaruRMP1982} This system has wide applications in the context of plasma physics and is of considerable interest from the fundamental point of view as an example of classical many-particle systems with extremely soft interactions.

In three dimensions (3D) the charges are interacting with each other via the unscreened Coulomb potential $V(r)=e^2/r$ and the OCP properties can be characterized by the single coupling parameter~\cite{BausPR1980,IchimaruRMP1982} $\Gamma=e^2/aT$. Here $e$ is the charge, $T$ is the temperature (in energy units), $a=(4\pi n/3)^{-1/3}$ is the 3D Wigner-Seitz radius, and $n$ is the particle density. In two dimensions (2D) the particles are interacting with the logarithmic potential~\cite{Caillol1982,Leeuw1982} $V(r)=-e^2\ln(r/a)$, which corresponds to the solution of the 2D Poisson equation and represents the interaction between infinite charged filaments (the two-dimensional system of particles with Coulomb interactions, which is also referred to as the 2D OCP, is not considered here).  
Here $a=(\pi n)^{-1/2}$ is the 2D Wigner-Seitz radius ($n$ being the 2D number density). The coupling parameter in the 2D OCP is~\cite{Caillol1982} $\Gamma=e^2/T$. Note that in the considered case it does not depend on density.

The collective modes of both 3D and 2D OCP have been extensively investigated.~\cite{BausPR1980,HansenPRL1974,HansenPRA1975,Leeuw1983,SchmidtPRE1997,GoldenPoP2000,
MithenAIP2012,KhrapakPoP2016_1,KhrapakPoP2016_2}
In the limit of weak coupling ($\Gamma\ll 1$) the Bohm-Gross dispersion relation applies,
\begin{equation}\label{BG}
\omega^2=\omega_{\rm p}^2+3k^2 v_{\rm T}^2,
\end{equation}
where $\omega$ is the frequency,  $k$ is the wave number, $\omega_{\rm p}=\sqrt{4\pi e^2n/m}$ (in 3D) and $\omega_{\rm p}=\sqrt{2\pi e^2n/m}$ (in 2D) is the plasma frequency, $m$ is the particle mass, and $v_{\rm T}=\sqrt{T/m}$ is the thermal velocity. When the coupling increases and the correlational effects set in ($\Gamma\gtrsim 1$), the dispersion relation (\ref{BG}) is substantially modified. One particular manifestation of the strong coupling effects is the {\it onset of negative dispersion}, referring to $d\omega/dk <0$ at $k\rightarrow 0$, which is in strong contrast with the prediction of the mean-field theory (\ref{BG}). 

The onset of negative dispersion was first discovered in early molecular dynamics simulations of the 3D OCP.~\cite{HansenPRL1974} The value of the coupling parameter at which the dispersion changes from positive to negative was initially estimated~\cite{HansenPRL1974,HansenPRA1975} as $\Gamma\gtrsim 3$. 
More recent molecular dynamics (MD) simulation, designed specifically to investigate the onset of negative dispersion,~\cite{MithenAIP2012} located the transition at $\Gamma\simeq 9.5$. Another set of MD simulations performed to acquire high quality structure function data for the classical OCP also allowed to follow positive-to-negative transition of the slope of the dispersion curve.~\cite{KorolovCPP2015} The borderline behaviour corresponded to $\Gamma\simeq 10$. 

The conventional approaches to the OCP dynamics (like for instance hydrodynamic treatment or quasi-localized charge aproximation) are incapable to describe accurately the onset of negative dispersion~\cite{BausPR1980} (see below for more details). More sophisticated theoretical models were therefore proposed. They demonstrated better agreement with the results from numerical simulations. For example, Carini {\it et al.}~\cite{CariniPLA1980} estimated the onset of negative dispersion to occur at $\Gamma\simeq 6$.  Later, Hansen~\cite{HansenJPL1981} proposed a model, which predicted the onset at $\Gamma\simeq 10$. In addition, he showed that the negative dispersion should also occur in the 2D  OCP with the onset at $\Gamma\simeq 14$.

Generally speaking, the onset of negative dispersion is a very stringent test of any theoretical model designed to describe accurately the collective dynamics of strongly coupled OCP fluid.
The purpose of this Brief Communication is to describe a rather simple model, which is able to capture the onset of negative dispersion in the OCP. The model is based on simple physical arguments and is applicable to both 3D and 2D cases. The predicted values of $\Gamma$ at which the onset occurs are in reasonable agreement with the results from simulations and more involved theoretical models. In the following,
the 3D OCP is first considered in detail. Generalizations to the 2D OCP are straightforward and will be discussed towards the end of the paper.      

It is convenient to start with the brief analysis of simple conventional tools used to describe the collective phenomena in OCP fluids. The standard hydrodynamic approach yields the dispersion relation of the form 
\begin{equation}\label{HD}
\omega^2=\omega_{\rm p}^2+\mu k^2 v_{\rm T}^2,
\end{equation}    
where $\mu=(1/T)(\partial P/\partial n)$ is the inverse reduced compressibility, which is also related to the bulk modulus $K$ via $\mu=(K/nT)$.
The onset of negative dispersion corresponds then to the coupling strength at which the inverse compressibility becomes negative. This value somewhat depends on the nature of the thermodynamic process. For instance, the isothermal compressibility of the OCP changes sign at $\Gamma\simeq 3$,~\cite{BausPR1980,MithenAIP2012} while the adiabatic compressibility becomes negative at $\Gamma\simeq 5$.~\cite{KhrapakPRE2014} 
Both these values are somewhat too low, compared to the results from numerical simulations.

The conventional quasi-localized charge approximation (QLCA) yields in the long-wavelength limit (to order $k^2$)~\cite{GoldenPoP2000}
\begin{equation}\label{QLCA1}
\omega^2=\omega_{\rm p}^2+\frac{4}{15}u_{\rm ex} k^2 v_{\rm T}^2,
\end{equation} 
where $u_{\rm ex}$ is the reduced excess energy (energy per particle in unuts of the system temperature) of the OCP. This dispersion is {\it always negative} since the excess energy is negative. However, it is well known~\cite{GoldenPoP2000,DonkoJPCM2008} that QLCA, as a theory for the strongly coupled state, neglects direct thermal effects. A simple phenomenological recipe for an extension~\cite{HouPRE2009} is to just add the corresponding kinetic term to Eq.~(\ref{QLCA1}). The dispersion becomes
\begin{equation}\label{QLCA2}
\omega^2=\omega_{\rm p}^2+k^2 v_{\rm T}^2\left(3+\frac{4}{15}u_{\rm ex}\right) ,
\end{equation}   
which coincides with the result based on the sum-rule analysis.~\cite{BausPR1980}
The reduced excess energy of the 3D OCP fluid can be accurately approximated with a simple three-term fit formula~\cite{KhrapakPoP2014,KhrapakCPP2016}
\begin{equation}\label{uEx}
u_{\rm ex} (\Gamma)=-0.9\Gamma+0.5944\Gamma^{1/3}-0.2786,
\end{equation} 
based on extensive Monte Carlo study by Caillol.~\cite{CaillolJCP1999} The onset of negative dispersion should occur then at $\Gamma\simeq 13.8$, close to the condition $\Gamma\geq 14$, quoted earlier.~\cite{BausPR1980} This, however, considerably exceeds the critical coupling parameter identified in numerical simulations.

Equations (\ref{HD}) and (\ref{QLCA2}) appear differently, but their structural composition is similar. Both contain the kinetic terms (``ideal gas'' contribution from uncorrelated particles) and the ``excess'' term associated with the correlational effects [the pressure and hence the inverse compressibility in Eq.~(\ref{HD}) can naturally be divided into ideal gas and excess contributions]. Moreover, the kinetic and excess terms appear in the dispersion relation as a simple superposition. This superposition is quite natural since the kinetic contribution dominates at weak coupling while the correlational effects produce dominant contribution at strong coupling. If the superposition is postulated, the form of the dispersion relation at long wavelengths can be readily guessed.   
The kinetic contribution should clearly be described by the Bohm-Gross terms (\ref{BG}). This ensures that the dispersion is exact when correlations are absent (weak coupling limit). The effect of correlations in the long-wavelength limit should be related to the bulk modulus, namely its {\it excess} component. Taking into account the character of the longitudinal (plasmon) mode dispersion of the OCP, the most physically relevant quantity is the high-frequency or instantaneous bulk modulus, $K_{\infty}$. The guessed long-wavelength dispersion relation of the OCP becomes 
\begin{equation}\label{Model}
\omega^2=\omega_{\rm p}^2+3k^2 v_{\rm T}^2+(\Delta K_{\infty}/nm)k^2.
\end{equation}
The arguments used to arrive at this expression are of course not rigorous, they are merely based on the physical intuition. The onset of negative dispersion is a very good opportunity to verify the validity of this simple construction. 

The high-frequency (instantaneous) elastic moduli of simple fluids were obtained by Zwanzig and Mountain~\cite{ZwanzigJCP1965}and Schofield.~\cite{Schofield1966} The usefulness of considering their excess constituent has been also discussed in the literature.~\cite{Stratt1997}  An expression for the excess contribution to the high-frequency bulk modulus $\Delta K_{\infty}$ can be derived using particularly simple route, which is briefly reproduced below. The starting point is the virial expression for the excess pressure,~\cite{HansenBook}
\begin{equation}\label{Pex}
P_{\rm ex}= -\frac{n^2}{6}\int_0^{\infty}rV'(r)g(r)d^3 r,
\end{equation}
where $V'(r)$ is the first derivative of the pairwise interaction potential and $g(r)$ is the radial distribution function (RDF). The excess bulk modulus is $\Delta K=-V(\partial P_{\rm ex}/\partial V)\equiv n(\partial P_{\rm ex}/\partial n)$, where $V$ is the volume containing $N$ particles, so that $n=N/V$. Differentiating (\ref{Pex}) gives
\begin{equation}\label{dK}
\Delta K= 2P_{\rm ex}-\frac{n^3}{6}\int_0^{\infty}rV'(r)\frac{\partial g(r)}{\partial n} d^3 r,
\end{equation}
where the second term on the right-hand side includes the implicit density dependence of the RDF. This is generally process-dependent. In the high frequency (instantaneous) limit no relaxation is allowed and the density change occurs without any rearrangement of the particles.~\cite{SingwiPRA1970} Mathematically, this implies 
\begin{displaymath}
\frac{d}{d n}g(r,n)=\frac{d}{d n}g(x n^{-1/3},n)=0,
\end{displaymath} 
or, equivalently,
\begin{equation}\label{RDF3D}
\frac{\partial g (r,n)}{\partial n}=\frac{r}{3n}\frac{\partial g(r,n)}{\partial r}.
\end{equation} 
Substituting this into Eq.~(\ref{dK}) and integrating by parts we obtain
\begin{equation}\label{K3D}
\Delta K_{\infty}^{3D}=-\frac{4\pi n^2}{9}\int_0^{\infty}dr r^3g(r)\left[V'(r)-(r/2)V''(r)\right].
\end{equation} 
This coincides with the expression for the excess bulk modulus obtained earlier using more thorough considerations.~\cite{ZwanzigJCP1965,Schofield1966} In the case of the OCP considered here one needs to account for the neutralizing background, which is done by substituting $g(r)\rightarrow h(r)=g(r)-1$ in Eqs.~(\ref{Pex})-(\ref{K3D}). Using the Coulomb interaction potential the excess bulk modulus can be directly related to the excess pressure. The result is rather simple in the OCP case: 
\begin{equation}\label{BM_3D}
\Delta K_{\infty}^{3D}=\frac{4}{3}P_{\rm ex}. 
\end{equation}   
Now we can substitute this into equation (\ref{Model}) and take into account that $P_{\rm ex}/nm=\tfrac{1}{3}v_{\rm T}^2u_{\rm ex}$ for the 3D OCP. The long-wavelength dispersion relation becomes
\begin{equation}\label{DR3D}
\omega^2=\omega_{\rm p}^2+k^2 v_{\rm T}^2\left(3+\frac{4}{9}u_{\rm ex}\right).
\end{equation} 
This immediately yields the condition for the onset of negative dispersion, $u_{\rm ex}(\Gamma)=27/4$.
With the help of the approximation (\ref{uEx}) we finally get $\Gamma\gtrsim 8.5$, in good agreement with numerical simulations~\cite{MithenAIP2012,KorolovCPP2015} and more sophisticated theoretical approaches.~\cite{HansenJPL1981}

The very same approach can be applied to the 2D OCP. The only difference is that the 2D excess bulk modulus should be used. The latter is obtained following the same lines as in the 3D case, starting from the 2D virial pressure expression
\begin{equation}\label{Pex2}
P_{\rm ex}= -\frac{n^2}{4}\int_0^{\infty}rV'(r)g(r)d^2 r.
\end{equation}   
Modifying Eq.~(\ref{RDF3D}) to describe instantaneous perturbations in the 2D case, the two-dimensional analogue of the Eq.~ (\ref{K3D}) can be readily obtained, 
\begin{equation}\label{K2D}
\Delta K_{\infty}^{2D}=-\frac{\pi n^2}{4}\int_0^{\infty}dr r^2g(r)\left[V'(r)-rV''(r)\right].
\end{equation} 
Applying this to the 2D OCP we have to change $g(r)$ to $h(r)$  (just as in the 3D case) and substitute the logarithmic interaction potential in the equations above.
This results in the following relation:
\begin{equation}\label{BM_2D}
\Delta K_{\infty}^{2D}=P_{\rm ex}. 
\end{equation}   
The excess pressure of the 2D OCP is known {\it exactly}, $P_{\rm ex}=-\tfrac{1}{4}nT\Gamma$. This equation of state immediately follows from the virial pressure equation combined with the charge neutrality condition.~\cite{Caillol1982} The long-wavelength part of the dispersion relation of the 2D OCP is
\begin{equation}\label{DR2D}
\omega^2=\omega_{\rm p}^2+k^2 v_{\rm T}^2\left(3-\Gamma/4\right).
\end{equation} 
The negative dispersion occurs for $\Gamma>12$, which is reasonably close to the condition derived previously by Hansen.~\cite{HansenJPL1981} 

To conclude, a simple approach to the long-wavelength behaviour of the plasmon mode in classical 2D and 3D OCP has been formulated. The approach is able to describe the onset of negative dispersion upon increasing coupling with a reasonably good accuracy. It demonstrates the important role that the excess instantaneous bulk modulus is playing in describing the longitudinal mode. This observation can be important not only within the scope of OCP, but also for other plasma-related systems such as complex or dusty plasmas.      

This work was supported by the A*MIDEX project (Nr.~ANR-11-IDEX-0001-02) funded by the French Government ``Investissements d'Avenir'' program managed by the French National Research Agency (ANR).

\bibliographystyle{aipnum4-1}
\bibliography{OCP_Disp}

\end{document}